\documentstyle[epsf,aps,preprint,floats]{revtex}
\def\Ii{ I_{\rm i} }
\def\If{ I_{\rm f} }
\def\dK{ {\mit\Delta K} }
\def\Ki{ K_{\rm i} }
\def\Kf{ K_{\rm f} }
\def\cM{ {\cal M} }
\def\braketb#1#2{\langle #1 | #2 \rangle}
\def\braketr#1#2#3{\langle #1 || #2 || #3 \rangle}

\tighten
\draft
\preprint{UMIST/Phys/TP/98-3}
\begin{document}

\title{Microscopic calculation of transition intensities for
vibrational bands and high-K isomers}

\author{Takashi~Nakatsukasa}
\address{Department of Physics, UMIST, P.O.Box 88, Manchester M60 1QD, UK}
\author{Yoshifumi~R~Shimizu}
\address{Department of Physics, Kyushu University, Fukuoka 812-8581, Japan}

\maketitle
\begin{abstract}
We investigate the effect of
the Coriolis coupling and the residual interactions
upon the inter-band transition rates
for the vibrational bands and
the decay of two-quasiparticle high-$K$ isomers.
\end{abstract}

\section{Introduction}

The Coriolis coupling in rotating nuclei
often leads to significant consequences.
The back-bending phenomenon is a typical example in which a nucleon pair
(Cooper pair) is broken by the Coriolis force.
In order to understand the effect of Coriolis coupling,
it is important to measure both the energy spectra and the inter-band
electromagnetic transition rates.
Theoretically,
the cranking model provides a powerful tool to describe numbers of
phenomena in rotating nuclei,
such as the angular momentum alignment, the pairing phase transition,
the superdeformation, etc.
However, the cranking model has a apparent disadvantage of
the semiclassical treatment of nuclear rotation,
and does not produce an angular momentum eigenstate.
Thus, for calculation of the inter-band transition rates,
one may need the angular momentum projection
which would be a heavy computational burden.

We have developed a new simple method to take account of quantum angular
momentum algebra in the cranking approach \cite{SN96}.
The method allows us to calculate matrix elements of
the intrinsic moments in the unified model and
to see, from the microscopic point of view,
how the Coriolis coupling affects the transition rates.
The formula for $M1$, $E1$ and $E2$ transitions is
recapitulated in section~\ref{sec: formula}.

The electromagnetic transitions between vibrational bands and the
ground-state band are a good testing ground of the application,
since the experimental data are available up to high spin.
This will be discussed in section~\ref{sec: vib_bands}.
The decay of high-$K$ isomers by the Coriolis mixing mechanism
would be another interesting example to see
the higher-order Coriolis coupling effect, which will be
discussed in section~\ref{sec: high-K}.

\section{Intensity relation formula}
\label{sec: formula}

We show only the final formula for the transition amplitudes.
See Ref. \cite{SN96} for details.

Since we will discuss only even-even nuclei in this paper,
we assume that the ground-state band has the $K=0$ and
$\dK=K_{\rm f}-K_{\rm i}\leq 0$ 
for transitions from the excited bands ($K_{\rm i},I_{\rm i}$)
to the ground-state band ($K_{\rm f}=0,I_{\rm f}$).
Within the lowest-order Coriolis coupling and the next order corrections,
the $\lambda$-pole ($\lambda=1,2$) transition
amplitudes are written in a form
\begin{eqnarray}
 \frac{\braketr{\Kf\If}{\cM(\lambda)}{\Ki\Ii}}{\sqrt{2\Ii +1}}
    &=& \sqrt{
   \frac{(\If - \Kf)!\,(\If + \Kf+n)!}
        {(\If - \Kf-n)!\,(\If + \Kf)!}} \,
    \braketb{\Ii \Ki \lambda \,(-\dK+n)}{\If \,(\Kf + n)}
  \nonumber \\
    &&\hspace{0.2\textwidth}
    \times  Q_t \left( 1 + q \, [\If (\If+1) - \Ii (\Ii+1)] \,
          \right) \ ,
\label{GIR}
\end{eqnarray}
where the forbiddenness
$n=0$ for the $K$-allowed transitions and
$n=|\Delta K| - \lambda > 0$ for the $K$-forbidden transitions.

The form of this formula is identical to the generalized intensity
relation in the unified model \cite{BM75}, in which
however there is no systematic method to
calculate the intrinsic parameters $Q_t$ and $q$.
We have shown that there is a correspondence between 
the $I$-expansion in the unified model and
the $\omega_{\rm rot}$-expansion in the cranking model,
and have given a general method to calculate these parameters
\cite{SN96}.

\begin{figure}[t]
\epsfxsize=\textwidth
\centerline{\epsffile{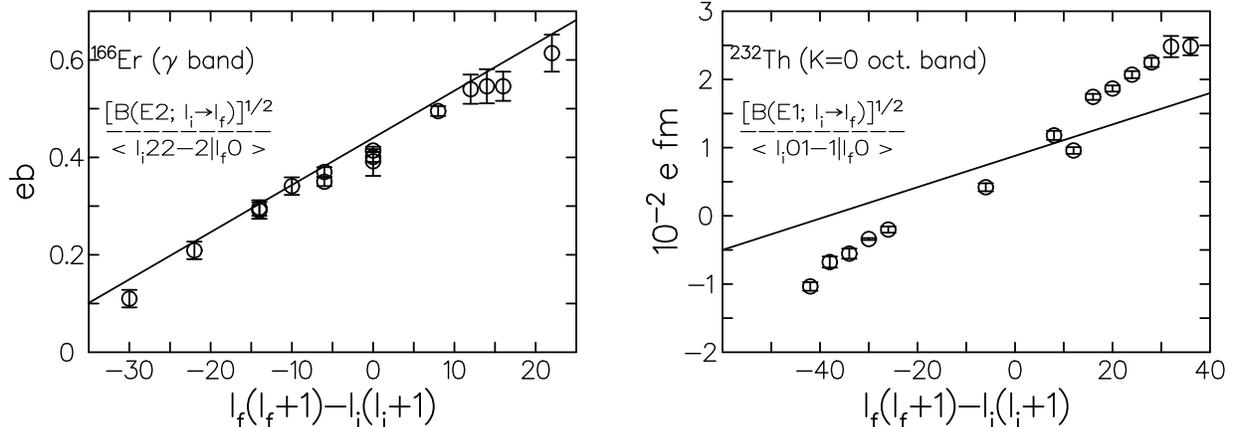}}
\caption{$E2$ transition amplitudes for $^{166}$Er (left) and
$E1$ transition amplitudes for $^{232}$Th (right).
The $E2$ data are taken from Fig.4-30 in Ref.\protect \cite{BM75}
and the $E1$ data are from Ref.\protect \cite{War99}.
The cranked RPA calculation has been carried out
using three major shells for protons and neutrons.
The calculated results are shown by solid lines.
}
\label{mikhailov}
\end{figure}
\section{Applications}
\subsection{$\gamma$- and octupole vibrational bands}
\label{sec: vib_bands}
The properties of vibrational motion in rotating nuclei may be
investigated by means of
a microscopic formalism of the random phase approximation (RPA) based on
the cranked mean field (See Ref. \cite{NMMS96} and references therein).
The method mentioned in the previous section is easily implemented to
the cranked RPA formalism.
Then, one can calculate the inter-band transition amplitudes
in a straightforward manner.
The calculation has been performed for quadrupole and octupole
vibrations to investigate $M1$, $E2$ and $E3$ transitions \cite{SN96}.
In Fig.~\ref{mikhailov},
we show $E2$ and $E1$ transitions for
for $\gamma$ vibration in $^{166}$Er and the $K=0$ octupole vibration
in $^{232}$Th.
For the $E2$ amplitudes of the $\gamma$ band,
the agreement with the experiments is almost perfect.
The $Q_t$ parameter for the $E1$ amplitude is difficult to reproduce
(the absolute value is renormalized in Fig.~\ref{mikhailov}),
while the Coriolis coupling parameter $q$ shows a reasonable agreement.

\begin{figure}
\epsfxsize=\textwidth
\centerline{\epsffile{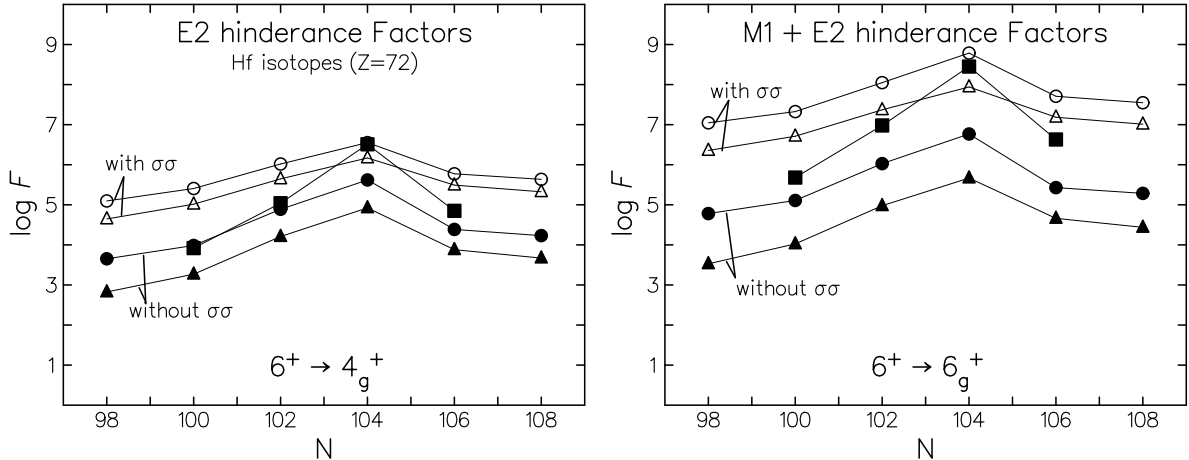}}
\caption{Hindrance factors of B(E2;$6^+\rightarrow 4_g^+$) (left)
and B(M1+E2;$6^+\rightarrow 6_g^+$) (right)
for $K^\pi=6^+$ isomers.
The squares are the experimental data, while
the circles indicate the calculated results.
The triangles are the same as circles but with
the proton chemical potential artificially increased by 70 keV.
The effective charge $e_\pi=2e$ and g-factor
$g_s^{\rm eff}=0.7g_s^{\rm free}$ have been used.
}
\label{hindrance}
\end{figure}
\subsection{Decay of High-$K$ isomers}
\label{sec: high-K}
If the nucleus has an axially symmetric shape,
the $K$ quantum number is supposed to be a good quantum number.
However, one can expect that this symmetry is broken in a strict sense
by the shape fluctuations and the Coriolis mixing.
The shape fluctuation with respect to the $\gamma$-degrees of freedom
has been studied by the $\gamma$-tunneling model \cite{NSS96}.
We discuss the Coriolis mixing mechanism for the
$K^\pi=6^+$ two-quasiparticle (2qp) isomers in Hf isotopes.

The configuration of the isomer is assumed to be the lowest
$K^\pi=6^+$ 2qp state in each nucleus which is actually
$\pi [402]5/2\otimes[404]7/2$.
The calculation involves the fourth (fifth) order of Coriolis coupling
for $E2$ ($M1$) decays,
and the hindrance factors are
estimated microscopically (filled symbols in Fig.~\ref{hindrance}).
The isotope dependence is qualitatively well reproduced,
however, the calculation predicts the hindrance factors too small.
Although the result is sensitive to the details of
quasiparticle spectra (see difference between
circles and triangles in the figure),
the decay rates are overestimated for most cases,
especially for the $M1$ transitions.

In order to investigate the effect of residual correlations,
we introduce the spin-spin interaction, $V_0{\bf\hat\sigma\hat\sigma}$,
which may be responsible for the Gallagher-Moszkowski splitting.
The interaction is diagonalized within a space of 2qp states
($E_{\rm 2qp}\leq 5$MeV),
with $V_0=100$ keV being adopted in this calculation.
The results are shown by open symbols in Fig.~\ref{hindrance}.
The hindrance factors are increased by 1$\sim$2 orders of magnitude
for $E2$ transitions
and by 2$\sim$3 orders for $M1$.
The results still depend on the details of quasiparticle spectra,
however, the dependence becomes weaker than the calculation
without the residual correlation.

\section{Conclusion}
In conclusion,
we have developed a general method to calculate the intrinsic
parameters in the intensity relations by using
the microscopic cranking approach.
The applications to the vibrational transitions and to decay of
high-$K$ isomers are discussed.
The calculated decay property of 2qp high-$K$ isomers seems
to be improved by the inclusion of the residual correlations.

\acknowledgments
This work is supported by a research grant from the Engineering
and Physical Sciences Research Council (EPSRC)  of Great Britain
(GR/L22331) and
by the Grant-in-Aid for Scientific Research from
the Japan Ministry of Education, Science and Culture (No. 10640275).

\end{document}